\documentclass[12pt]{iopart}

\usepackage{graphicx}
\begin{document}

\title[Control of the ferroelectricity in metastable state for orthorhombic $R$MnO$_{3}$ crystals]{Control of the ferroelectricity in metastable state for orthorhombic $R$MnO$_{3}$ crystals}

\author{K. Noda, M. Akaki, F. Nakamura, D. Akahoshi, \\and H. Kuwahara}

\address{ Department of Physics, Sophia University 
Chiyoda-ku, Tokyo 102-8554, JAPAN}
\ead{n-kohei@sophia.ac.jp}
\begin{abstract}
We have investigated orthorhombic $R$MnO$_{3}$ ($R$=(Gd$_{1-y}$Tb$_{y}$)) crystals near the phase boundary between the paraelectric $A$-type-antiferromagnetic (AF) phase (PA) and the ferroelectric transverse-spiral-AF one (FS). 
The spiral AF structure breaks inversion symmetry and induces the ferroelectric polarization through the inverse Dzyaloshinskii-Moriya (DM) interaction. 
We have found that the PA-FS phase boundary is located at around 0.15$<$$y$$<$0.2. 
In $y$=0.15 compound, PA and FS phases appear in cooling scan, while FS is not observed in warming scan. 
This result suggests that FS observed in $y$=0.15 is a metastable state arising from the competition between PA and FS.
Furthermore, we demonstrate the phase control between these competing phases by the application of the external magnetic field and external quasihydrostatic pressure.
\end{abstract}


\section{Introduction}

Multiferroic materials are now the subject of renewed interest.
Orthorhombic $R$MnO$_{3}$ ($R$=rare earth ions) containing TbMnO$_{3}$ famous as the ``magnetic control of the ferroelectric polarization" \cite{kimura} is one of such multiferroic materials, and some of them are simultaneously ferroelectric and antiferromagnetic. 
In $R$MnO$_{3}$, the magnetic frustration gives rise to the change of magnetic structures (canted $A$-type AF phase $\Rightarrow$ transverse spiral AF one $\Rightarrow$ $E$-type AF one) as a function of the ionic radius of $R$, and such magnetic frustration comes from the competition between the nearest neighbor (NN) ferromagnetic and the next NN antiferromagnetic (AF) interactions \cite{Kimura-2}

TbMnO$_{3}$ and DyMnO$_{3}$ show the transverse-spiral-AF phase, and in such magnetic phase the spontaneous ferroelectric polarization is observed \cite{Kenzelmann,Arima-2}. 
Recently, it is considered that such noncollinear-transverse-spiral magnetic-structure breaks inversion symmetry and induces the ferroelectric polarization through the inverse Dzyaloshinskii-Moriya (DM) interaction \cite{Kenzelmann,Arima-2,Katsura,Mostovoy,Dagotto}.
On the other hand, the canted-$A$-type-AF phase, which is observed in $R$MnO$_{3}$ with a larger ionic-radius than $R$=Gd, does not induced the ferroelectric polarization, and the anticorrelation between paraelectric canted-$A$-type-AF phase (PA) and ferroelectric transverse-spiral-AF phase (FS) is reported \cite{goto-2}. 
Therefore, the PA and FS phases compete with each other between GdMnO$_{3}$ and TbMnO$_{3}$. 

In this work, we have investigated the PA-FS phase boundary in $R$MnO$_{3}$ ($R$=Gd$_{1-y}$Tb$_{y}$) crystals in external magnetic field and under external quasihydrostatic pressure, to clarify the magnetoelectric properties of the phase boundary and to control the phase transition between PA and FS.

\section{Experiments}
All single crystals used in this work were prepared by the floating zone method (in Ar gas at a pressure of 0.25MPa).
We carried out an x-ray diffraction experiment to confirm that all crystals had an orthorhombic $Pbnm$ structure without any impurity phase. 
All specimens used in this study were cut along the crystallographic principal axes into a rectangular shape by means of an x-ray back-reflection Laue technique. 
Measurements of the temperature dependence of the dielectric constant and the spontaneous ferroelectric polarization in magnetic fields were carried out in a temperature-controllable cryostat equipped with a superconducting magnet that provides a field up to 8T\@.
The dielectric constant was determined with an LCR meter (Agilent, 4284A). 
After the sample had been cooled under a poling electric field of 300$\sim$500 kV/m, the spontaneous electric polarization was obtained by the accumulation of a pyroelectric current while it was heated at a rate of 4K/min. 
An external quasihydrostatic pressure was produced by a clamp-type piston cylinder cell using Fluorinert as the pressure-transmitting medium.

\section{Results and Discussion}

\subsection{Phase control through the chemical pressure}

\begin{figure}[ht]
\begin{center}
\includegraphics[clip,scale=0.48]{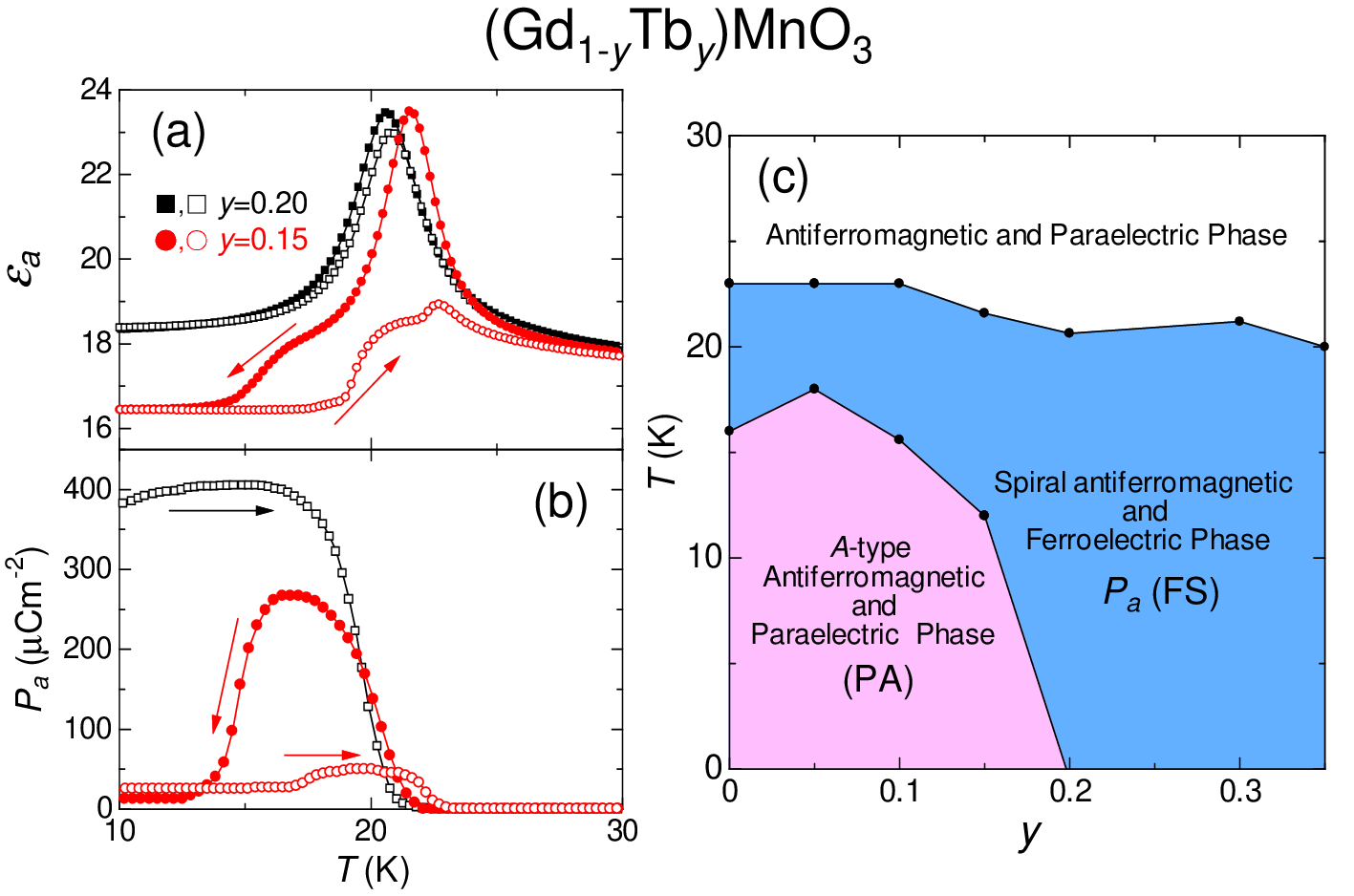}
\vspace{-3mm}
\caption{The temperature dependence of the dielectric constant along the $a$ axis ($\varepsilon_{a}$) (a) and ferroelectric polarization along the $a$ axis ($P_{a}$) (b) of (Gd$_{1-y}$Tb$_{y}$)MnO$_{3}$ ($y$=0.15 and 0.20). Solid and open symbols display a cooling and warming scan respectively. Solid arrows indicate scan directions. (c) the magnetic and electric phase diagram of (Gd$_{1-y}$Tb$_{y}$)MnO$_{3}$ (0 $\leq$ $y$ $\leq$ 0.35) obtained from cooling scan without a magnetic field. }
\label{fig1}
\end{center}
\end{figure}

At first, we have investigated the PA-FS phase boundary by the control of the chemical composition of $R$MnO$_{3}$ ($R$=(Gd$_{1-y}$Tb$_{y}$)). 
As a result, we have found that the phase boundary exists at around 0.15$<$$y$$<$0.2. 
Figures \ref{fig1} (a) and (b) show the temperature dependence of the dielectric constant along the $a$ axis ($\varepsilon_{a}$) and the spontaneous ferroelectric polarization along the $a$ axis ($P_{a}$) of $R$MnO$_{3}$ near the phase boundary ($y$=0.15 and 0.20). 

In the case of $y$=0.20, only FS is observed below the paraelectric-AF--FS transition temperature in both cooling and warming scan. (See Fig. \ref{fig1} (a) and (b)) 
$\varepsilon_{a}$ shows the divergent peak at 22K, below which $P_{a}$ appears. 
On the other hand, in the case of $y$=0.15, the reentrant (paraelectric-AF $\Rightarrow$ FS $\Rightarrow$ PA) behavior is observed. 
In the cooling scan, $\varepsilon_{a}$ shows the divergent peak at 21K and at this temperature $P_{a}$ appears. 
However, $P_{a}$ suddenly decreases below 16K and finally disappears at 13K.
$\varepsilon_{a}$ decreases with a decrease of $P_{a}$, and the value of $\varepsilon_{a}$ in the ground state takes a small value compared with that of $y$=0.20. 
This disappearance of $P_{a}$ is originated from the magnetic phase transition from the transverse-spiral-AF phase to the canted-$A$-type AF one. 
This disappearance of $P_{a}$ agrees well with the scenario of the anticorrelation between PA and FS \cite{goto-2}.
In warming scan, the FS is veiled by PA. (See Fig. \ref{fig1} (a) and (b)) 
Therefore, FS in cooling scan is considered to be the metastable state. 
Such reentrant and hysteric behavior (or metastable state) is observed in $y$=0$\sim$0.15, and the FS-PA transition temperature decreases with an increase of $y$. (See Fig. \ref{fig1} (c))
These results suggest that such reentrant behavior arises from the competition between PA and FS.

\subsection{Phase control by the external magnetic field}

\begin{figure}[ht]
\begin{center}
\includegraphics[clip,scale=0.48]{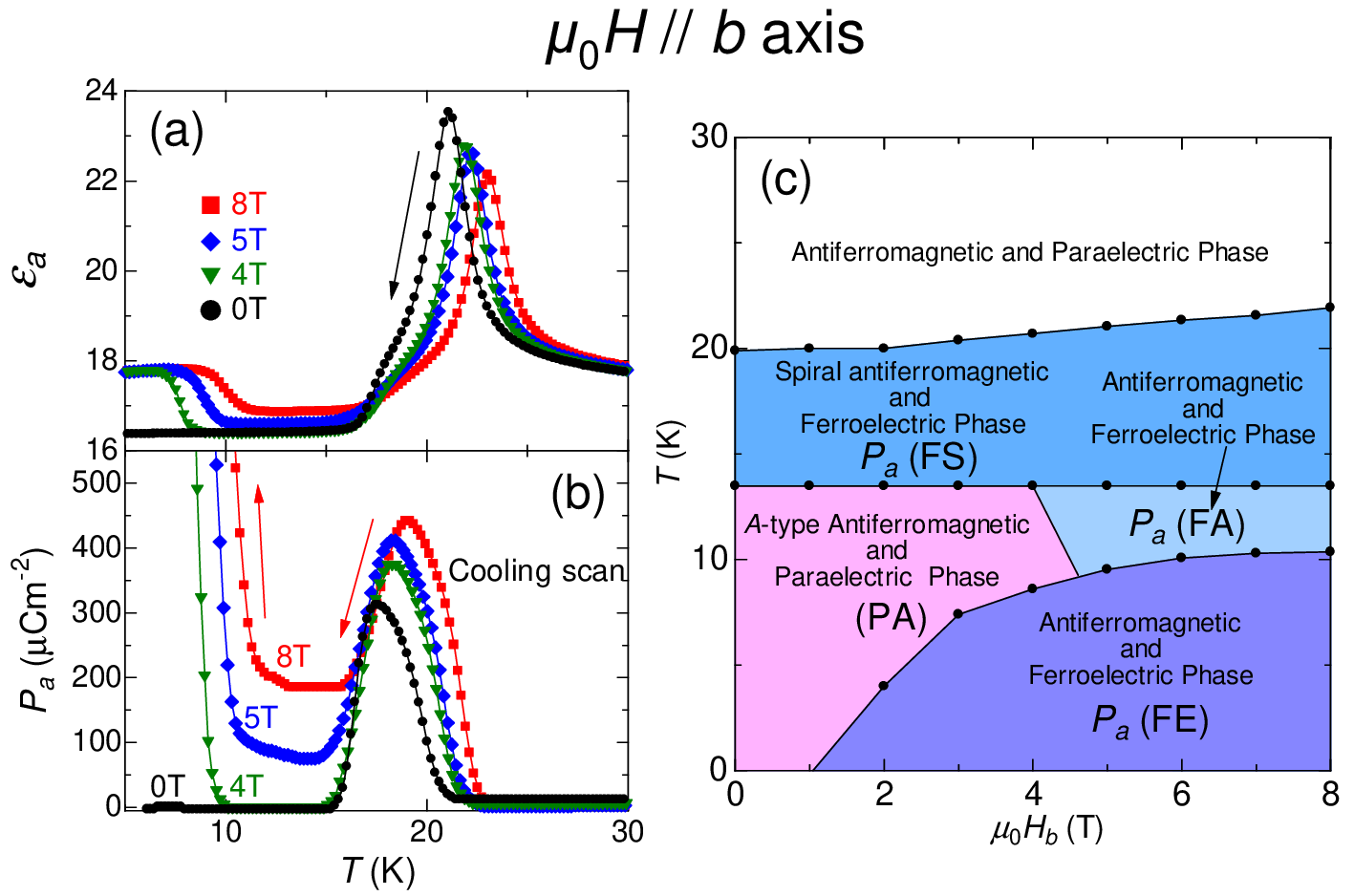}
\vspace{-3mm}
\caption{ The temperature dependence of dielectric constant along the $a$ axis ($\varepsilon_{a}$) (a) and ferroelectric polarization along the $a$ axis ($P_{a}$) (b) of (Gd$_{1-y}$Tb$_{y}$)MnO$_{3}$ ($y$=0.15) in the external magnetic field parallel to the $b$ axis ($H_{b}$) in. Only the cooling scan data is displayed. (c) the magnetic and electric phase diagram of (Gd$_{1-y}$Tb$_{y}$)MnO$_{3}$ ($y$-0.15) in $H_{b}$ in cooling scan. }
\label{fig2}
\end{center}
\end{figure}

Then, we demonstrate the phase control between PA and FS by the application of external magnetic field. 
The drastic phase conversion from FS to PA is realized by the application of the magnetic field parallel to the $c$ axis ($H_{c}$) \cite{goto-2}. 
Therefore, in this work, we try to convert PA to FS by the application of the magnetic field. 
Figures \ref{fig2} (a) and (b) show the temperature dependence of $\varepsilon_{a}$ and $P_{a}$ in the magnetic field parallel to the $b$ axis ($H_{b}$) in $y$=0.15. 

In $H_{b}$ $\leq$ 4T, the reentrant ferroelectric transition is observed and at this transition temperature the dielectric anomaly appears, while such reentrant transition and dielectric anomaly do not occur in a zero magnetic field. (See fig. \ref{fig2} (a) and (b))
The ferroelectric phase, which is observed in higher temperatures, is FS. 
On the other hand, the ferroelectric phase (FE), which appears by the application of $H_{b}$ in a lower temperature, is considered to arise from the magnetic order of Gd 4$f$ moments, and such ferroelectric phase has already reported in previous works (\cite{Kuwahara, Noda, Noda2}). (However, the magnetic structure of FE has not yet been clarified). 
These ferroelectric phases are separated by PA, and this result suggests that these ferroelectric phases have no relation to each other.  

Then, in $H_{b}$ $\geq$ 5T, PA changes to the ferroelectric phase (FA) by the application of $H_{b}$. 
The $P_{a}$ emerges by the application of $H_{b}$ and the value of $\varepsilon_{a}$ is becoming large gradually with the increase of $H_{b}$. (See the temperature region between 10K and 15K in fig. \ref{fig2} (a) and (b))
However, the decrease of $P_{a}$ is observed at around the PA-FS transition temperature in $H_{b}$ $\leq$ 4T. 
These results suggest that, in FA, the strong competition between PA and FS is induced by the application of $H_{b}$ above 5T, and such competition seems to cause the coexistence of PA and FS, while the anticorrelation between PA and FS is observed and such coexistence does not occur in $H_{c}$ \cite{goto-2}. 

The electromagnetic phase diagram of $y$=0.15 in $H_{b}$ is shown in Fig. \ref{fig2} (c). 
FE appears above 2T, and the ferroelectric transition temperature increases by the application of $H$$_{b}$ below 10K. 
FS is observed above 13K. 
The ferroelectric transition temperature of FS increases by the application of $H_{b}$, while the transition temperature, at which the $P_{a}$ disappears or decreases, is constant. 
Between these ferroelectric phases, FA is observed in $H_{b}$ above 5T. 
The influence of the 4$f$ moments seems to exist below 10K, because FE is observed below 10K. 
Therefore, the ferroelectric phases observed above 10K are considered to be originated from the magnetic structure of Mn 3$d$ spins only. 
So, in FA, the coexistence of PA and FS is ascribed to the competition between the spiral AF and canted $A$-type AF phases of Mn 3$d$ spins. 
These results suggest that FS is enhanced by the application of $H_{b}$, while PA is suppressed. 
This effect of $H_{b}$ is same as the effect of increase of $y$ in Fig \ref{fig1} (c).

\subsection{Phase control by the external pressure}

\begin{figure}[ht]
\begin{center}
\includegraphics[clip,scale=0.48]{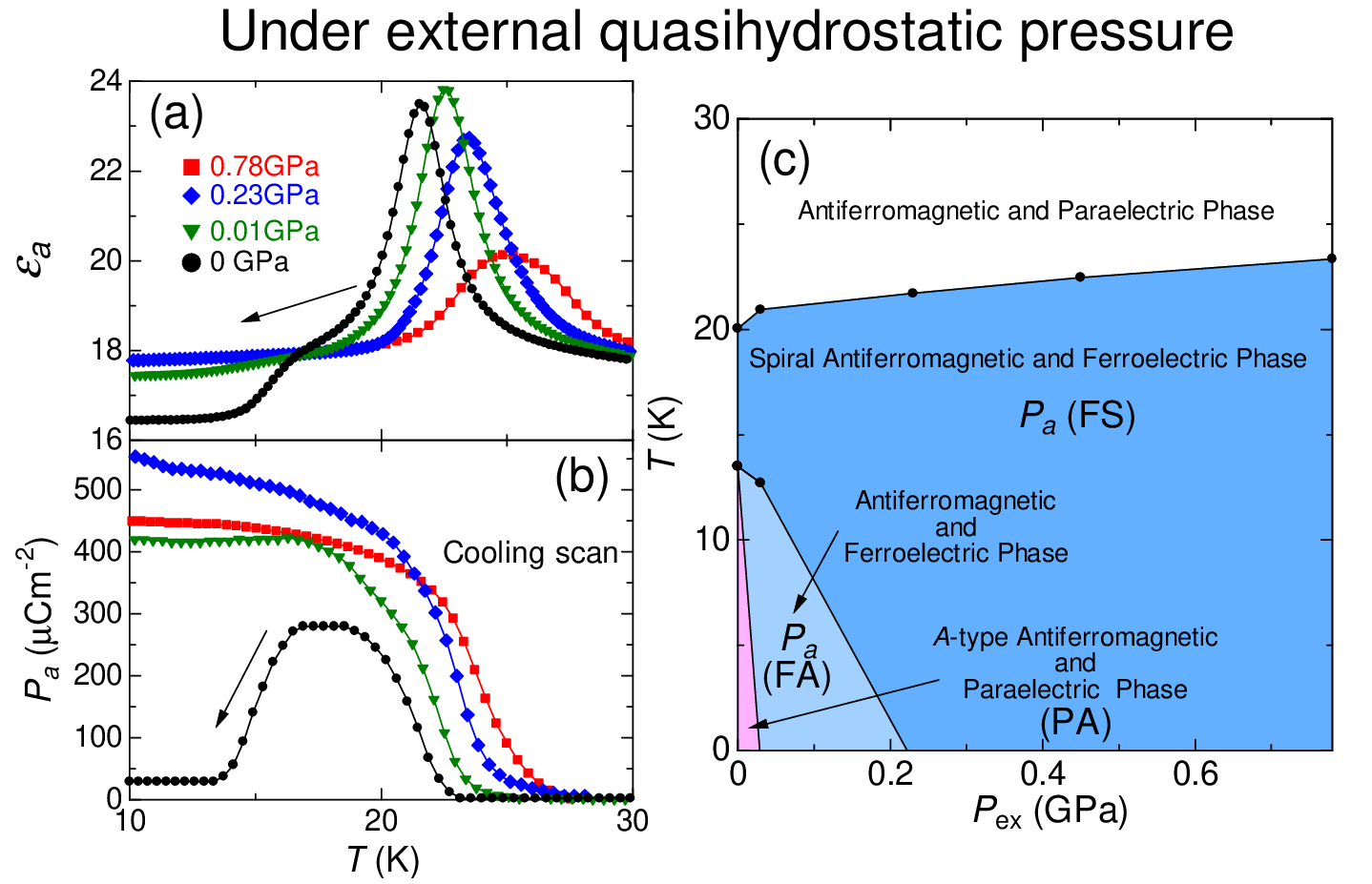}
\vspace{-3mm}
\caption{ The temperature dependence of the dielectric constant along the $a$ axis ($\varepsilon_{a}$) (a) and ferroelectric polarization along the $a$ axis ($P_{a}$) (b) of (Gd$_{1-y}$Tb$_{y}$)MnO$_{3}$ ($y$=0.15) in the external quasihydrostatic pressure ($P_{\rm ex}$). Only the cooling scan data is displayed. (c) the magnetic and electric phase diagram of (Gd$_{1-y}$Tb$_{y}$)MnO$_{3}$ ($y$=0.15) under $P_{\rm ex}$ in cooling scan. }
\label{fig3}
\end{center}
\end{figure}

In this section, we discuss the effect of the external quasihydrostatic pressure ($P_{\rm ex}$) on the magnetoelectric properties in orthorhombic $R$MnO$_{3}$ crystals. 
In LaMnO$_{3}$, at room temperature, the linear decrease of the lattice parameters with an increase of $P_{\rm ex}$ is reported \cite{Loa}. 
Therefore, the effect of $P_{\rm ex}$ on the magnetoelectric phases is considered to be same as the effect of the increase of $y$ in Fig \ref{fig1} (c). 

Figures \ref{fig3} (a) and (b) show the temperature dependence of $\varepsilon_{a}$ and $P_{a}$ under $P_{\rm ex}$ respectively. 
As shown in Figs. \ref{fig3} (a) and (b), PA easily collapses into FS with an increase of $P_{\rm ex}$. 
The value of $\varepsilon_{a}$ in the ground state becomes large with an increase of $P_{\rm ex}$. 
At 0.01GPa, the value of $\varepsilon_{a}$ is a little smaller than others, and the small decrease of $P_{a}$ is observed. 
This result suggests that the coexistence of PA and FS occurs under 0.01GPa. (Main phase is FS, but PA remains as minor phase. Therefore, this phase is considered to be same as FA observed in $H_{b}$.)

The electromagnetic phase diagram under $P_{\rm ex}$ is shown in Fig. \ref{fig3} (c). 
As shown in Fig. \ref{fig3} (c), PA is drastically transected into FS by the application of $P_{\rm ex}$. 
As expected, this result shows the same behavior as the increase of $y$ in Fig \ref{fig1} (c). 
This result agrees well with the scenario of the anticorrelation between PA and FS \cite{goto-2}, and indicates that the transverse-spiral-AF phase, which induces the ferroelectric polarization, arises from the magnetic frustration due to the increase of orthorhombic distortion.

\section{Summary}
In summary, we have studied the dielectric and magnetic properties of a (Gd$_{1-y}$Tb$_{y}$)MnO$_{3}$ single crystals. 
We have found the PA-FS phase boundary at around 0.15$<$$y$$<$0.2 in a zero magnetic field. 
In addition, we have observed the reentrant (paraelectric phase $\Rightarrow$ ferroelectric one $\Rightarrow$ paraelectric one) phase transition in the cooling scan. 
Such reentrant behavior probably arises from the strong competition between PA and FS. 
The phase control is achieved by the application of $H_{b}$ and $P_{ex}$. 
In contrast to the case of the application of $H_{c}$ \cite{goto-2}, the coexistence between PA and FS is induced by the application of $H_{b}$. 
In such coexisting phase, partial $P_{a}$ is observed; however the microscopic origin of this ferroelectric polarization has not yet been clarified. 
These results suggest that $H_{b}$ enhances FS and suppresses PA in $R$MnO$_{3}$. 
On the other hand, the drastic phase transition from PA to FS occurs by the application of $P_{\rm ex}$. 
Therefore, the magnetic frustration due to the increase of the orthorhombic distortion is considered to be the most important origin of the drastic phase transition, which provides an improved understanding of the mechanism of the magnetoelectric effect in multiferroic materials. 

\section*{Acknowledgement}
This work was supported by Grant-in-Aid for Scientific Research (C) from the Japan Society for Promotion of Science, and by Grant-in-Aid for Scientific Research on Priority Areas gHigh Field Spin Science in 100Th(No.451) from the Ministry of Education, Culture, Sports, Science and Technology (MEXT)D

\end{document}